\documentclass[12pt,preprint]{aastex}
\begin{document}
\title{Hypervelocity A \& B Stars should be slow rotators}
\author{Brad M. S. Hansen\altaffilmark{1} 
}
\altaffiltext{1}{Department of Physics \& Astronomy, and Institute of Geophysics \& Planetary Physics, University of California Los Angeles, Los Angeles, CA 90095, hansen@astro.ucla.edu}


\lefthead{Hansen }
\righthead{Hyper--V A \& B}

\begin{abstract}
The most commonly accepted explanation for the origin of hypervelocity stars in the halo of the Milky Way is that they are
the result of tidal disruption of binaries by the massive black hole at the center of the Galaxy. We show that, if this scenario
is correct, and if the original binary properties are similar to those in the local stellar neighbourhood, then the hypervelocity stars should rotate with velocities measureably lower than those for field stars of similar spectral type. This may prove to be a more direct test of the model than trying to predict the position and velocity distributions.
\end{abstract}


\section{Introduction}

The discovery of a population of young, high velocity stars in the halo of the
Galaxy (Brown et al. 2005), termed hypervelocity stars, is quite naturally explained in terms of ejection
from the region around the massive black hole at the center (e.g. Hills 1988). 
The original proposal by Hills was that stellar binaries on grazing orbits are
tidally disrupted by the black hole, leaving one star bound and one unbound with a 
high ejection velocity. The connection is further enhanced by the fact that the
hypervelocity stars are late-B type stars and thus of similar spectral type to the
population of peculiarly young stars present in the central parsec (Ghez et al. 2003, 
Eisenhauer et al. 2005), although 
this is partly the result of sample selection (Brown et al. 2006). 

Examining the dynamical link between the hypervelocity stars and the so-called Sag~A$^*$
cluster may help to explain the puzzling origins of this latter population. The 
modern realisation of Hills' scenario invokes the scattering (Perets, Hopman \& Alexander 2007) of young stellar binaries
onto nearly radial orbits by massive perturbers (usually giant molecular clouds) from Galactocentric
radii $\sim $ several parsec (where such stars can form more naturally than in the
strong tidal field of the central black hole). The dynamical disruption of these binaries then
produces the observed hypervelocity stars and a bound population that results in the
Sag~A$^*$ cluster (Gould \& Quillen 2003, Ginsburg \& Loeb 2006). However, some
issues remain. The apastra of the cluster stars do not extend as far out as the Hills scenario
would suggest and the scenario does
not naturally explain the second population of young stars in the Galactic center, the
so-called He-emission line stars (e.g. Paumard et al. 2006), as these stars are too far from the black hole to
be the result of tidal disruption. 

At least two alternative scenarios exist for the origin of the hypervelocity stars.
They may also be ejected through two-body encounters with massive, dark objects
orbiting the central black hole (Yu \& Tremaine 2003). Indeed, the existence of a second, intermediate mass, black
hole ($\sim 10^3$--$10^4 M_{\odot}$) in the Galactic center has been suggested (Hansen \& Milosavljevic 2003)
as a mechanism for explaining the presence of the young stars as the tidally stripped remnants of
an inspiralling stellar cluster (Gerhard et al. 2001). In this scenario there is no direct one-to-one
link between hypervelocity stars and Sag~A$^*$ cluster stars, although both are assumed to be drawn from the same
parent population.

In a related fashion, stars may be ejected by two-body encounters with stellar mass black 
holes $\sim 10 M_{\odot}$ (O'Leary \& Loeb 2006), a population of which is expected to form a cusp
around the central black hole as a result of mass segregation in the Galactic potential (Morris 1993;
Miralda-Escude' \& Gould 2003). The efficiency of this process is lower than scattering by an
IMBH, and  more subject to stellar destruction by tidal disruption or physical collisions (Ginsburg \& Loeb 2007). However,
it may still be significant and may be applicable if the observed young stars had their
origins in a gravitationally unstable accretion disk (Levin \& Beloborodov 2003) rather than a tidally
disrupted cluster.

The original scenario due to Hill requires a binary, whereas the scenarios in which stars scatter off
black holes do not. 
Furthermore, the orbital separations of the binaries that give rise to hypervelocity stars are small
enough (Hills 1988; Gualandris, Portegies Zwart \& Sipior 2005; Bromley et al. 2006) that the stellar rotation of the star is affected by the
rotation of hypervelocity stars to test whether they did indeed spend some part of their lifetime
in a binary. If the rotation velocities of hypervelocity stars are similar to those of binary stars of the same
spectral type,
 and measureably different to equivalent single stars, then we may interpret
this as evidence in favour of the binary disruption origin of hypervelocity stars.

In \S~\ref{Bin} we examine the empirical trends of stellar rotation with binary separation, for early-type
stars, and how this translates into a relation between stellar rotation and ejection velocity, via the
auspices of  Hills' scenario. Observational confirmation would provide a strong argument in favour of
this model, as the alternative models do not require binary stars and thus do not imply the same
consequence.

\section{A relation between rotation and ejection velocity}
\label{Bin}

The rotational properties of early-type stars in binaries have been compiled by Abt \& Boonyarak (2004).
Their most striking finding is that stars in binaries seem to show a measureable decrease in rotation
velocity (relative to single stars of similar type) at separations well beyond the limit at which tides
are expected to operate. For B-type stars, the most relevant to the observations of Brown et al., the
trend of $V \sin i$ with $P_{orb}$ is flat out to $P_{orb} \sim 10^4$~days, with a mean velocity
of $83 \pm 16$~km/s, measurably lower than the value of $134 \pm 7$ km/s for single stars (Abt, Levato \& Grosso 2002).

More powerful discriminants are possible if the observational selection is extended to encompass stars of
early A~type. For stars of spectral type A0-A5, Abt \& Boonyarak show that the rotation of stars in the
 shortest period binaries show the expected synchronisation trend, but that 
significantly reduced rotation extends out
as far as orbital periods $\sim 100$~days, before returning to rates consistent with single stars at longer 
periods. An orbital period of 100~days corresponds to a separation of 0.77~AU (for a binary consisting of
two 3$M_{\odot}$ main sequence A stars), which yields an ejection velocity of 935~km/s using equation~1 from
Bromley et al. (2006). Thus, the orbital separations of interest for hypervelocity stars are precisely those
for which rotation velocities are affected by the presence of a companion.

\subsection{Velocity-Velocity relation}

\begin{figure}
\plotone{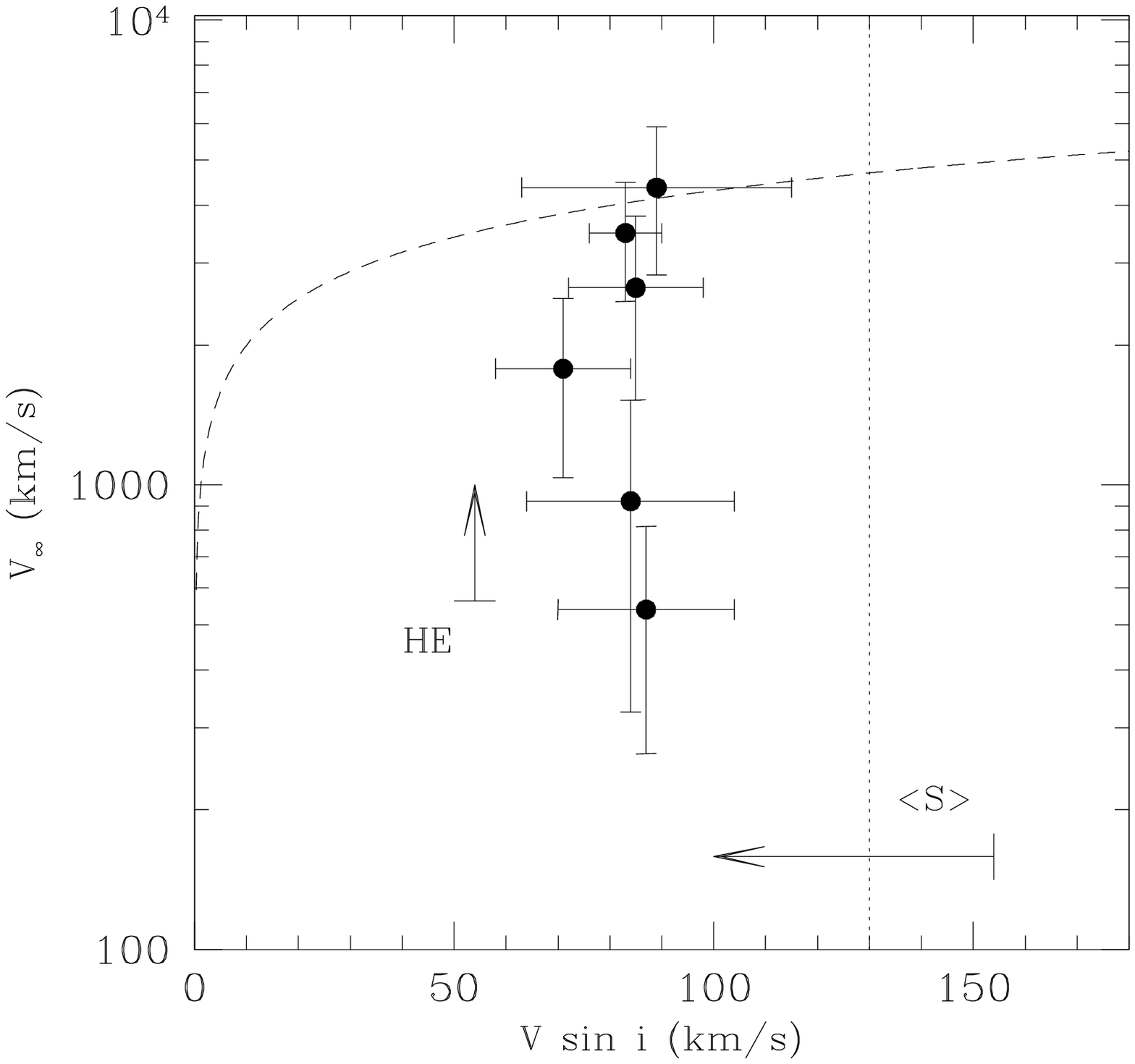}
\figcaption[f1.ps]{ The points show the relationship between rotation velocity and ejection
velocity expected, on the basis of Hills' scenario, for B-type stars. The vertical dotted line indicates the average
rotation velocity of single stars. The dashed line assumes strict synchronisation of the star with the
original binary rotation, with an ejection velocity given by equation~1 of Bromley et al (2006) and
randomly oriented inclinations.
The arrow at the bottom right is the upper limit on the average rotation velocity
of the Sag A$^*$ cluster B-type stars from Eisenhauer et al. (2005). The arrow at left shows the parameters of the star HE~0437-5439 (Edelmann et al. 2005), where we have used the measured radial velocity as
a lower limit, since it has presumably slowed as it climbed out of the Galactic potential well.
 \label{vvB}}
\end{figure}

\begin{figure}
\plotone{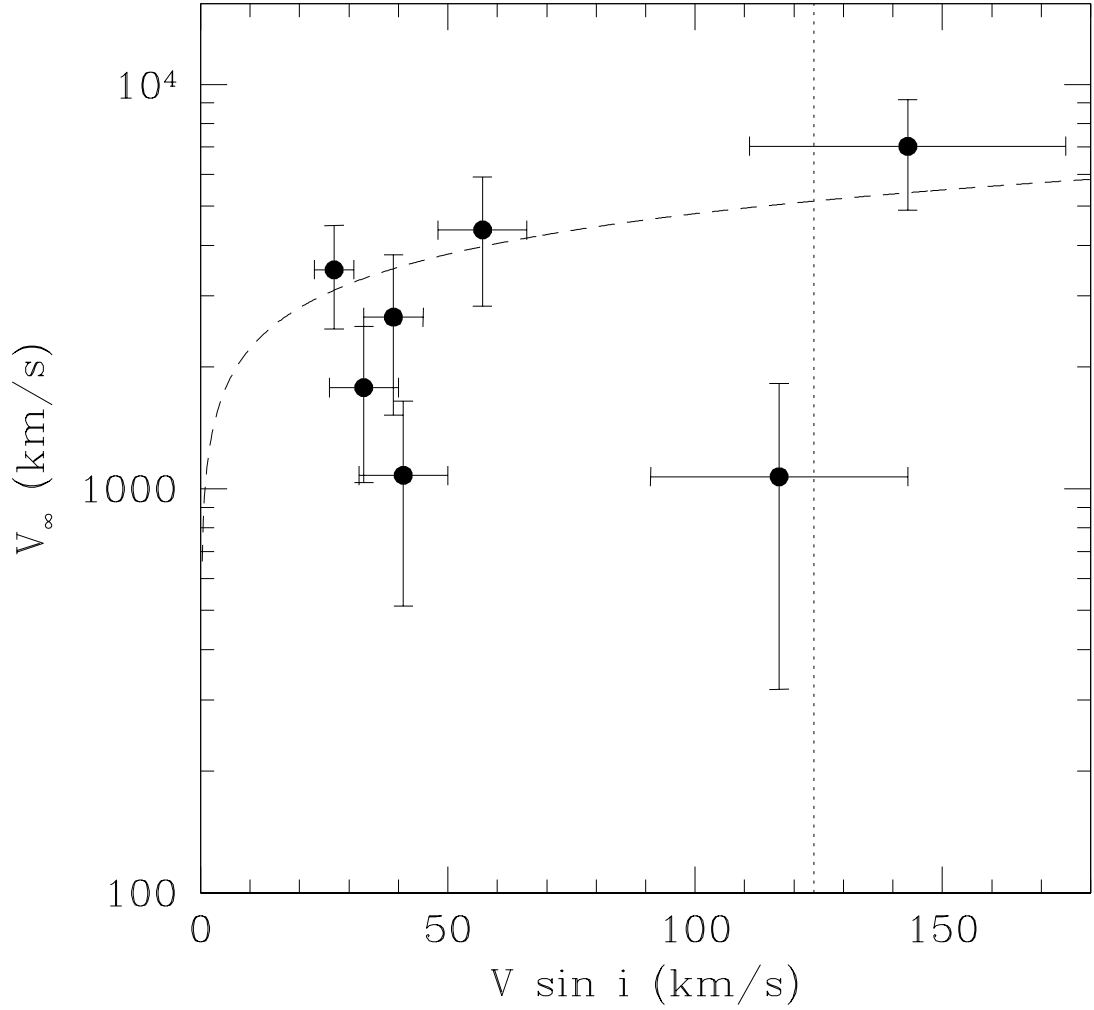}
\figcaption[f2.ps]{ The points show the expected relationship between rotation velocity and ejection
velocity, on the basis of Hills' scenario, for early A stars. The vertical dotted line indicates the average
rotation velocity of single stars and the dashed line is as described in Figure~1.
 \label{vv}}
\end{figure}

In order to investigate how the rotational properties of the original binary members are related to their
final fates as hypervelocity stars, we have performed our own version of the scattering calculations of
Hills, Gualandris et al. and Bromley et al., using a Bulirsch-Stoer integrator with energy precision
$\epsilon = 10^{-9}$ for each step-size. We allow for random binary orientations and assume a logarithmic
distribution of initial separations, for a range from 0.01 to 10~AU. We start the binaries at a distance of
5000~AU from the center, with
velocities of 250 km/s and a range of impact parameters, weighted by equal probability per unit area, out to
impact parameters of 3300~AU (which corresponds to closest approach distances of $\sim 100$AU). As noted by
previous authors, there is a distinct trend indicating that higher ejection velocities result from the disruption
of closer binaries. However, there is significant scatter in the relation, as a result of the other variables
in the problem, mostly related to the orientation of the binary relative to its trajectory.
 For each of the initial period bins used by Abt \& Bonnyarek, we average over the
distribution of final $V_{\infty}$, to derive the relationship between average  $V \sin i$ and 
average $V_{\infty}$ (emerging from the Galactic center).
 The resulting relationship
is shown in 
Figure~\ref{vvB} for B-stars and in Figure~\ref{vv} for A0-A5 stars. We see that the hypervelocity B stars are expected to be systematically
slower rotators than the $\sim 130$ km/s measured for isolated B stars (Abt et al. 2002). For the A stars, there is more structure
in the relation. At the highest ejection velocities (resulting from the closest binaries), the spins follow
the synchronisation trend, but there is a low rotation tail that extends down to velocities $\sim 1000$ km/s, before the rotation returns to values more characteristic of single stars.
In both figures, the error bars on $V_{\infty}$ indicates the dispersion about the mean, rather than the
error in the mean, as in the case of $V \sin i$.

One might wonder whether the  passage of the star close to the central black hole can have an effect on
the spin. However, the distances of closest approach are rarely close enough ($<$~AU) 
for there to be a significant tidal spin-up effect, especially since there is only one passage past the
black hole. For the member of the original binary that remains bound, repeated encounters with other stars
may have an effect over time (Alexander \& Kumar 2001), but the ejected star should remain unaffected.

What of those stars that remain bound to the black hole?
Ghez et al. (2003) report a rotation of $220\pm 40$ km/s for the star SO-2, which suggests that this star
is not a natural fit to the favoured scenario. This is corroborated by the fact that the measured eccentricity
(e=0.87), while high, is not high enough ($e >0.97$) to match the expectations of the simple binary disruption model 
(as noted in Ginsburg \& Loeb 2006, and which our simulations confirm).
For the other S-stars, Eisenhauer et al. (2005) find an upper limit to the average rotation velocity of the Sag A$^*$ cluster
stars of 154 km/s, by co-adding the spectra. This shows they are, at least, not significantly spun-up by
tidal interactions, either in very close binaries or by the central black hole.

\subsection{Passage into the Halo}

We must also account for the deceleration of the ejected stars as they climb out of the potential
well to their present observed positions.
To answer this, we generate specific
realisations, following the approach of Bromley et al. We sample the distribution of ejected velocities
shown above directly, integrating the orbits outward in the Galactic potential model of Bromley et al., with
values $\alpha=2$, $a_c=8 pc$ and $\rho_0=1.27 \times 10^4 M_{\odot}.pc^{-3}$. We integrate for a random
value of the age between 0 and the turnoff age. For 3$M_{\odot}$, we take this to be 350~Myr and
for 4$M_{\odot}$ we adopt 160~Myr. We select
stars that lie between 10~kpc and 100~kpc (the approximate observational volume). 

\begin{figure}
\plotone{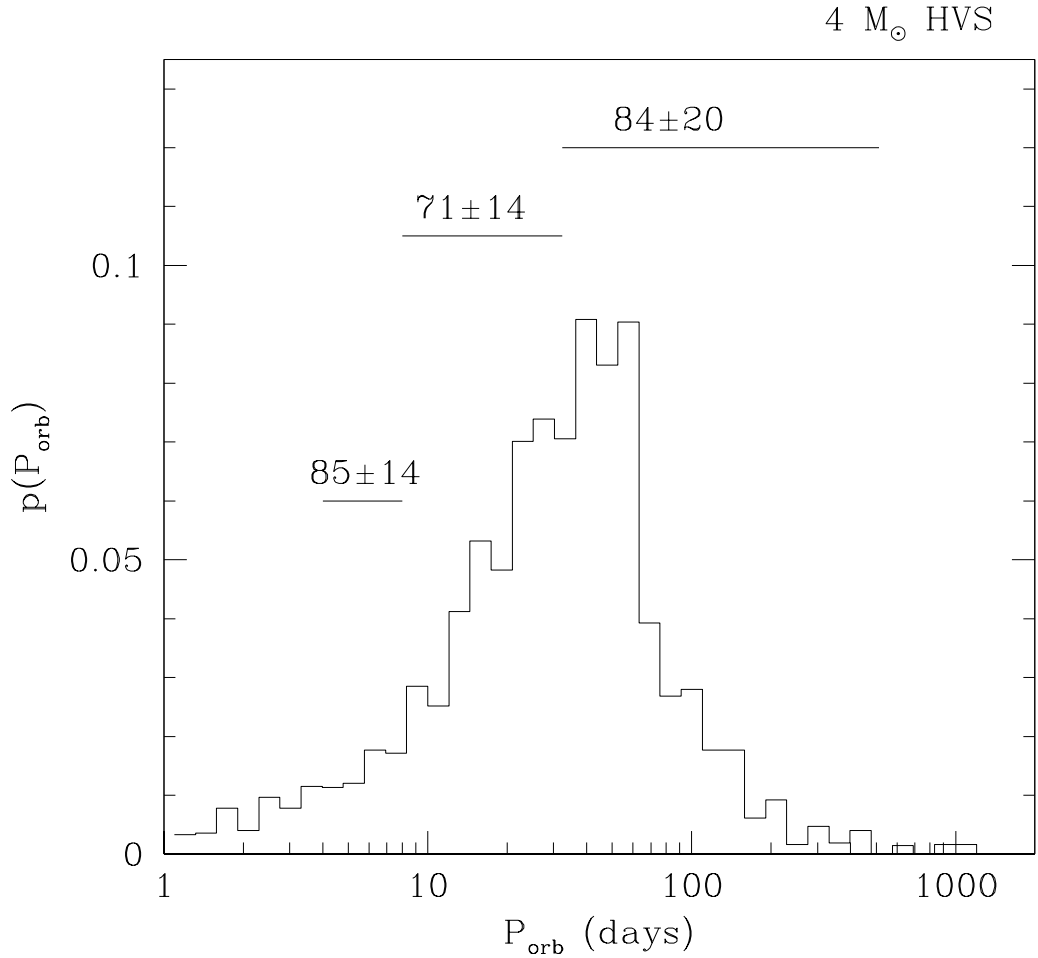}
\figcaption[f3.ps]{ Histogram of the original binary orbital periods for the observed hypervelocity B stars.
Also indicated are the mean rotation velocities measured by Abt \& Boonyarek for field binaries in each
of the designated period ranges. We see that the bulk of the hypervelocity stars should be slow rotators.
 \label{NPB}}
\end{figure}

Figure~\ref{NPB} shows the distribution of original orbital periods
of the binaries that give rise to the observable hypervelocity B stars.
 Interestingly, we see a strong peak
at $\sim 40$--$60$~days and that the vast majority of the observed stars had original orbital periods $<100$~days. Thus, we can make the strong prediction that the observed rotation velocities of the stars observed by Brown et al. should be $\sim 70-90$km/s.

\begin{figure}
\plotone{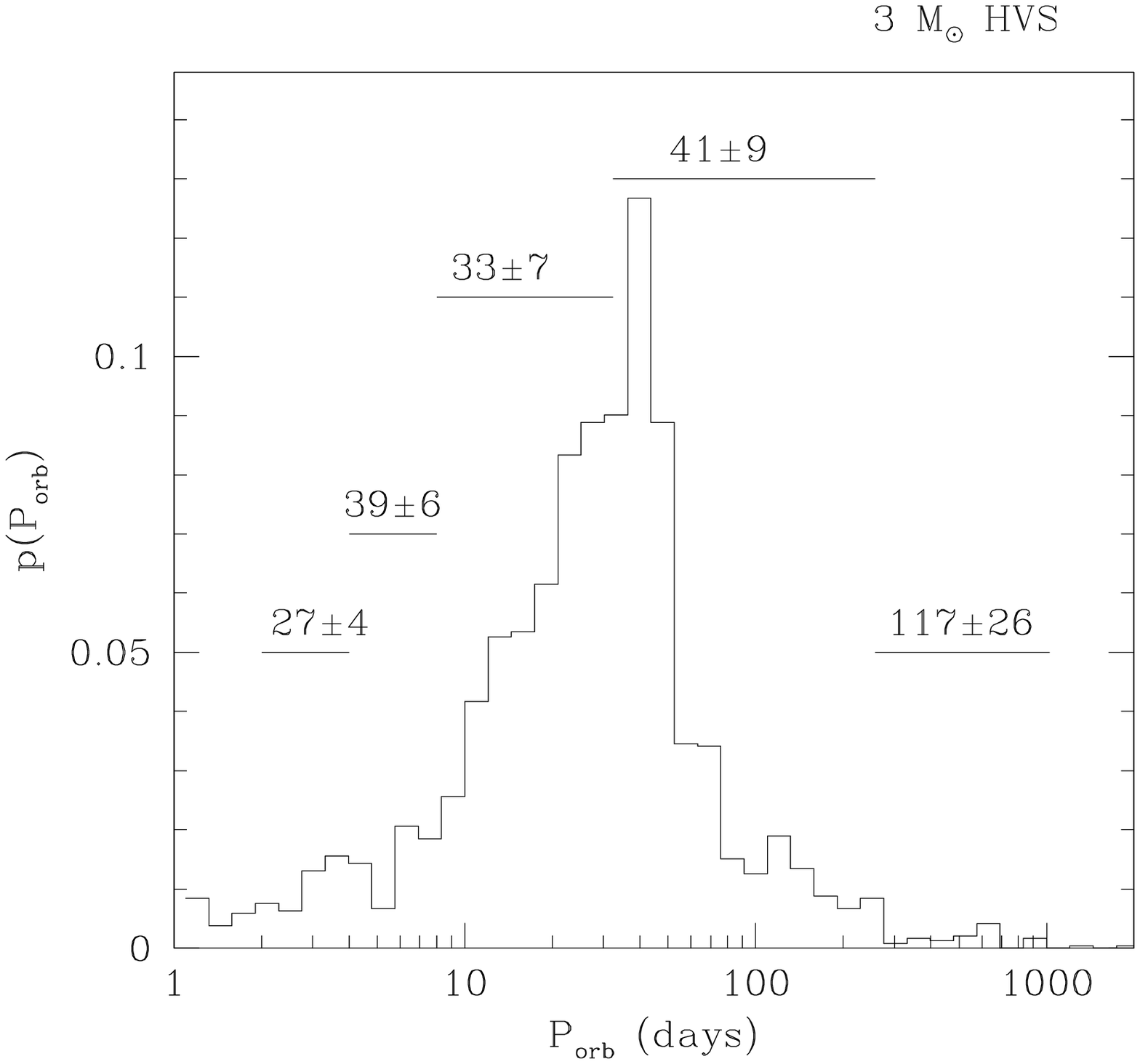}
\figcaption[f4.ps]{ Histogram of the probability distribution of the original binary orbital periods for hypervelocity A stars in the
observational volume.
Also indicated are the mean rotation velocities measured by Abt \& Boonyarek for field binaries in each
of the designated period ranges. We see that the bulk of the hypervelocity stars should be slow rotators.
 \label{NPA}}
\end{figure}

Figure~\ref{NPA} shows the same distribution in the case of $3 M_{\odot}$, which we take to
represent the early A stars. 
We see that we can make
an even stronger prediction regarding the spins of hypervelocity early A stars, namely that the great
majority should be very slowly
rotating, with $V \sin i \sim$ 30--40 km/s. Stars with rotation velocities $>100$~km/s (common among single A stars)
 should be 
quite rare amongst the hypervelocity stars.

\section{Discussion}

The currently most widely accepted model for the origin of hypervelocity stars is that they correspond to
the break-up of binaries that interact with the massive black hole at the Galactic center. We have
shown that, based on the empirically determined properties of early-type stars in binaries, and
the kinematic requirements of the ejection scenario,  one should expect the hypervelocity stars to
rotate slower than the average for single stars of their type. This is the result of the fact that, in
order for a star to attain 
the velocities observed, it must come from  a system whose initial orbital separation is 
small enough that the rotation is affected by the presence of a binary companion.

In quantitative terms, 
we have shown that the rotation velocities for the observed late-type B stars are expected to be
$\sim 80$km/s, compared with $\sim 130$km/s one finds for most single B-type stars. If the observational
selection can be pushed into the regime of the early-type A stars, the prediction gets even stronger,
with expected rotation velocities $\sim 40$km/s. Furthermore, it is worth noting that
 the discovery of hypervelocity Am stars would
be a direct signature of initial binarity, since the peculiar spectral signatures of Am stars are the
result of radiative levitation in atmospheres that are very quiescent due to their slow rotation. This is
why Am stars are almost always found in binaries. (Abt 1965, 2000)

The preceding assertions assume that the properties of the initial binaries are accurately reflected
by the studies of those in the solar neighbourhood. However, this is implicit in  Hills' scenario,
in that
it invokes stars which formed far enough from the black hole that there is no reason to expect the
star formation properties to be any different (unlike many scenarios for the origins of the young stars
in the central parsec, where the physics is likely quite different). Another underlying assumption that
one might question is whether the binaries have had enough time to circularise before being disrupted
(i.e. are the solar neighbourhood binaries, in some sense, older)? To answer this, we note Abt \& Boonyarek's
empirical result that the regime of quasi-synchronisation extends to orbital periods well beyond that expected
by simple tidal theory. Thus, it is likely the result of processes at work during the formation of the binary, and which then imprint the rotational properties of the stars well before the binary is disrupted. Even in
the case where the synchronisation is the result of classical tidal interaction, for early type stars the
tidal forces are quite sensitive to the stellar structure, and most of the tidal synchronization
takes place in the early stages of the star's evolution (Zahn 1977).

Thus, we consider it a robust prediction that hypervelocity stars resulting from Hills' scenario should
be slowly rotating, and measureably different from the rotational properties of field stars of similar
spectral type.
If hypervelocity stars are instead found to be rotating like normal field B or A stars, then this would favour
 one of the other ejection scenarios, such as scattering by an intermediate mass black hole (Levin 2006,
Baumgardt, Gualandris \& Portegies Zwart 2006; Sesana, Haardt \& Madau 2006). Such a measurement should be within reach of current generations of telescopes. Indeed, the spectra of Brown et al. have a velocity 
resolution of $\sim 100$ km/s, so that only a modest increase of a factor of 2 in spectral resolution is
required to start putting some real constraints on the ejection scenario. In fact, the hypervelocity star
HE~0437-5439 (Edelmann et al. 2005), already has a measured rotation velocity of $V \sin i = 54 \pm 4$ km/s. This low value is nicely consistent with the expectations of the binary disruption scenario, although its possible origin in the LMC is a complicating factor.

In almost all cases, if one binary member is ejected, the other remains bound to the black hole in
a highly eccentric orbit. Thus, the above considerations apply to these stars as well. Indeed, Hills'
scenario has been invoked to explain the origin of the so-called Sag~A$*$ cluster. However, the best-studied
member of this group, the star SO-2, does not agree well with the predicted properties from the Hills'
scenario. The eccentricity is not large enough to be the result of a binary disruption, and it is rotating
normally, in contrast to the above prediction. In principle, 
the presence of other (possibly dark) bodies in the same vicinity may lead to additional perturbations which
might explain the present observed parameters. However, the short age of the star renders this explanation
problematical. As always, we await further observation.

\acknowledgements 
The author acknowledges useful comments from Tal Alexander, Andrea Ghez, Hagai Perets and Simon
Portegies Zwart.


\newpage

\newpage

\newpage

\begin{references}
\reference{Abt65} Abt, H. A.,  ApJS, 11, 429
\reference{Abt00} Abt, H. A., ApJ, 544, 933
\reference{AB04} Abt, H. A. \& Boonyarak, C., 2004, ApJ, 616, 562
\reference{ALG} Abt, H. A., Levato, H. \& Grosso, M., 202, ApJ, 573, 359
\reference{AK} Alexander, T. \& Kumar, P., 2001, ApJ, 549, 948
\reference{BK} Bromley, B. C. et al., 2006, ApJ, 653, 1194
\reference{B05} Brown, W. R., Geller, M. J., Kenyon, S. J., \& Kurtz, M. J., 2005, ApJ, 622, L33
\reference{B05} Brown, W. R., Geller, M. J., Kenyon, S. J., \& Kurtz, M. J., 2006, ApJ, 647, 303
\reference{BGP} Baumgardt, H., Gualandris, A. \& Portegies Zwart, S., 2006, MNRAS, 372, 174
\reference{ENH} Edelmann, H., Napiwotzki, R., Heber, U., Christlieb, N. \& Reimers, D., 2005, ApJ, 634, L181
\reference{E05} Eisenhauer, F., et al., 2005, ApJ, 628, 246
\reference{G01} Gerhard, O., 2001, ApJ, 546, L39
\reference{G03} Ghez, A. M. et al., 2003, ApJ, 586, L127
\reference{GL} Ginsburg, I. \& Loeb, A., 2006, MNRAS, 368, 221
\reference{GL07} Ginsburg, I. \& Loeb, A., 2007, MNRAS, 376, 492
\reference{GQ} Gould, A. \& Quillen, A. C., 2003, ApJ, 592, 935
\reference{GPPZ} Gualandris, A., Portegies Zwart, S. \& Sipior, M. S., 2005, MNRAS, 363, 223
\reference{H88} Hills, J. G., 1988, Nature, 331, 687
\reference{LB03} Levin, Y. \& Beloborodov, A., 2003, ApJ, 590, L33
\reference{L06} Levin, Y., 2006, ApJ, 653, 1203
\reference{MG} Miralda-Escude', J. \& Gould, A., 2000, ApJ, 545, 847
\reference{M93} Morris, M., 1993, ApJ, 408, 496
\reference{OL} O'Leary, R. M. \& Loeb, A., 2006, astro-ph/0609046
\reference{P06} Paumard, T., et al., 2006, ApJ, 643, 1011
\reference{PHA} Perets, H. B., Hopman, C. \& Alexander, T., 2007, ApJ, 656, 709
\reference{SMH} Sesana, A., Haardt, F. \& Madau, P., 2006, ApJ, 651, 392
\reference{YT} Yu, Q. \& Tremaine, S., 2003, ApJ, 599, 1129
\reference{Z77} Zahn, J.-P., 1977, A\&A, 57, 383
\end{references}
\end{document}